\documentclass{article} 		

\usepackage{amsmath}
\usepackage{amsfonts}
\usepackage{amssymb}
\usepackage[UKenglish]{babel}
\usepackage{graphicx}

\usepackage{mathptmx}       
\usepackage{helvet}         
\usepackage{courier}        
\usepackage{type1cm}        
\usepackage{makeidx}         
\usepackage{graphicx}        
\usepackage{multicol}        
\usepackage[bottom]{footmisc}

\makeindex             

\newtheorem{theorem}{Theorem}[section]

\begin{document}

\title{Tensor Product Approximation (DMRG) and Coupled Cluster method in Quantum Chemistry}


\author{\"Ors Legeza, Thorsten Rohwedder, Reinhold Schneider and Szil\'ard Szalay}

\date{}



\maketitle

\abstract{ We  present the Copupled Cluster (CC) method and the Density matrix Renormalization Grooup  (DMRG)  method in a unified way,
 from the perspective of recent developments in tensor product approximation. 
We present an introduction into recently developed {\em hierarchical tensor representations}, in particular {\em tensor trains} which are
matrix product states in physics language. The discrete equations of  full CI approximation applied to  the electronic Schr\"odinger equation is casted into 
 a tensorial framework in  form of the second quantization. A further approximation is  performed afterwards by tensor approximation within a hierarchical format or
 equivalently a tree tensor network. We establish the  (differential) geometry of low rank hierarchical tensors and apply the Driac Frenkel principle to 
 reduce the original  high-dimensional problem to low dimensions. The DMRG algorithm is established as an optimization method  in this format 
 with alternating directional search.  We briefly introduce the CC method and  refer to our  theoretical results.
 We compare this approach in the present discrete formulation  with the  CC method and its underlying exponential parametrization.   
}

\section{Introduction}
\label{sec:Intro}

The Coupled Cluster (CC) method has been established during the past two decades 
as a standard approach for computing the electronic structure of molecules 
whenever high accuracy is required and attainable \cite{bart_CC_overv,helgaker}. 
Density Functional Theory (DFT) is still suffering from modeling errors,
however, due to low scaling complexity 
it allows the treatment of relatively large systems. Among the several chapters dedicated to DFT in this book, please see
in particular the chapter of Tzanov and Tuckerman and the chapter Watermann et al. and the chapter of Ghiringhelli
for a more detailed discussion about computational advantages and limitations of current DFT methods. 
Recent improvement of DFT models in order to obtain more accurate results on one hand, 
and low order scaling techniques for CC to get rid of the computational burden, on the other hand,
have brought both approaches to similar limitations. 
According to this development CC is no longer restricted to small systems,
and due to its superior accuracy, has gained increasing  interest for practical applications.  

On the other hand, both methods are applicable  only to systems which can be approximated appropriately by single particle models. 
This situation is often referred to as \emph{dynamical} or \emph{weak correlation}. 
CC using a restricted Hartree-Fock (HF) determinant 
can describe the ground state of a closed-shell molecule nearly up to  basis set error.
Perhaps, it fails whenever this determinant is insufficient to describe the physics qualitatively. 
For example if the closed-shell molecule separates into open-shell molecules,
or if the ground state is nearly degenerated.
In this situation only multi-reference representations are appropriate. 
Since there is no precise distinction between weak and strong correlation,
or dynamical and static  correlation, 
it explains roughly how to distinguish between nice and hard problems. 

The Density Matrix Renormalization Group (DMRG) algorithm and Matrix Product States (MPS) 
are more recent techniques which seem to be based on a completely different perspective.
Here correlation is replaced by \emph{entanglement}. 
When a system is decomposed into interacting subsystems,
entanglement describes the quantum correlation among them.
This approach is already established for the computation of  quantum lattice systems, 
like spin systems or the Hubbard model \cite{white,schollwoeck},  
but DMRG is less well established in quantum chemistry.

The present article  presents the CC method and the DMRG method in a unified way,
namely from the perspective of recent developments in tensor product approximation \cite{kolda,belmol,hackbusch}. 
In the traditional framework,
tensor product approximation has provided the fundamentals of quantum chemistry,
namely Hartree Fock as anti-symmetric rank-one approximation 
and variational multi-configurational methods 
like Multi-Configurational Self-Consistent Field (MCSCF), 
or Multi-Configuration Time Dependent Hartree (MCTDH) and quantum dynamics \cite{helgaker}.
Although we are starting from the electronic Schr\"odinger equation, 
we will take a basic knowledge about these methods for granted.  
The extended tensor framework has evolved hidden in the renormalization group ideas, 
and became clear in the framework of matrix product and tensor network states. 
Independently of these developments, it has been introduced in quantum dynamics 
as the multi-layer MCTDH method \cite{mayer,wang}, 
and recently in numerics as hierarchical tensor representation, 
namely, Hierarchical Tucker (HT) \cite{hackbusch,hackbuschkuehn} 
and Tensor Trains (TT) \cite{osele,oselea}. 

Since we have not found the material in an elementary form in the literature,
we have designed this article in a tutorial style. 
To keep the presentation short and compact,
the present article is not considered as a complete survey article,
with an extended bibliography and historical remarks. 
We have also omitted advanced techniques,
like explicitly correlated CC methods or low-order scaling techniques,
e.g.~we refer to \cite{schuetz-werner,klopper,densityfitting}.  
The multi-reference CC method as a topic of intense current research is also beyond the scope of this present presentation. 
We refer  the readers 
to the excellent recent survey articles \cite{schollwoeck,densityfitting,chan} 
and the monograph \cite{helgaker}.


\section{Electronic Schr\"odinger Equation and Second Quantization}
\label{sec:SchSecQuant}

There are many different notations in the literature.
For convenience, here we list our conventions in advance.
In the following, $N$ denotes the number of electrons,
and $d$ is the dimension of the one-particle Hilbert space.
We use $i$, $j$, $a$, $b$, $p$ or $q$ for the indexing of \emph{orbitals},
which are then running in $1,2,\dots d$,
while $\xi$ or $\zeta$ are used for the indexing of \emph{particles},
they are then running in $1,2,\dots N$.
Greek letters $\mu$ or $\nu$ stand for \emph{occupation numbers}, taking vaules in $0,1$ for fermions.
These correspond to $x$ in the general language of tensor network description,
(where they are running in $1,2,\dots,n_i$)
applied to the second quantized formalism,
in which framework $k$ is also used for internal bond indices (where they are running in $1,2,\dots,r_i$).
(Note that indexed indices of the form $p_\xi$, or $\mu_i$, $x_i$, $k_i$ make sense.)
Greek letters $\alpha$ or $\beta$ stand for \emph{indices of excitation operators} in the CC ansatz.

\subsection{Electronic Schr\"odinger equation}
\label{subsec:SchSecQuant.Sch}

We will describe two alternative approaches, the DMRG and the CC methods,
to solve the \emph{stationary electronic Schr\"{o}dinger equation} numerically,
by approximating the exact wave-function $\Psi$. 
The \emph{electronic Schr\"{o}dinger equation} describes 
the stationary behaviour of a non-relativistic quantum mechanical system
of $N$ electrons in a field of $K$ classical nuclei of charge $Z_{\eta}\in\mathbb{N}$
and fixed positions $\mathbf{R}_{\eta}\in\mathbb{R}^{3}$. 
It is an operator eigenvalue equation for the Hamiltonian $H$ of the system, given by
\begin{equation*}
 H= -\frac{1}{2}\,\sum_{\xi=1}^N \Delta_\xi 
    - \sum_{\xi=1}^{N}  \sum_{\eta=1}^K\frac {Z_\eta}{| \mathbf{r}_\xi-\mathbf{R}_\eta|} 
  + \frac{1}{2}\sum_{\xi=1}^{N}\sum_{\zeta=1\atop \zeta \neq \xi}^{N}  \frac{1}{| \mathbf{r}_\xi-\mathbf{r}_\zeta|},
\end{equation*}
which acts on \emph{wave functions} $\Psi$ that depend on 
$N$ spatial coordinates $\mathbf{r}_{\xi}\in\mathbb{R}^{3}$ and 
$N$ spin coordinates $s_{\xi}=\pm\frac{1}{2} \in \mathbb{Z}_2$ of the $N$ electrons. 
The \emph{Pauli principle} requires the  wave functions $\Psi$ to be antisymmetric w.r.t.~the particle variables. 
This means that $\Psi$ changes sign under permutation of two distinct variable pairs $(\mathbf{r}_{\xi}, s_\xi) \leftrightarrow (\mathbf{r}_\zeta,s_\zeta)$. 
The energy space  of $H$, i.e.~the space of wave functions, is 
\begin{equation*}
 \mathbb{H}^1_N   =    H^1 \Bigl(\mathbb{R}^{3}\times 
 \mathbb{Z}_2, \mathbb{C}\Bigr)^{N} 
\cap \bigwedge_{\xi=1}^N L_2 \Bigl(\mathbb{R}^{3}\times 
\mathbb{Z}_2, \mathbb{C}\Bigr),
\end{equation*}
with $H^1(X, \mathbb{K})$ denoting the set of $\mathbb{K}$-valued weakly differentiable functions on $X$, 
ans the symbol $\wedge$ is used for the antisymmetric tensor product of spaces.
Due to well known regularity results $\Psi$ has a certain {Sobolev regularity}, see e.g.~\cite{yserlect}. 
For ground state computation it is sufficient to consider only real valued functions.
There $\mathbb{K}=\mathbb{C}$ could be replaced by $\mathbb{R}$.  
In its variational, or weak formulation \cite{yserlect},
the electronic Schr\"odinger equation consists of finding $\Psi \in \mathbb{H}^1_N$ 
and an eigenvalue $E^*\in \mathbb{R}$ such that 
\begin{equation}
 \langle \Phi, H\Psi \rangle  =  E^* \langle \Phi, \Psi \rangle, \quad \text{for all}\; \Phi \in \mathbb{H}^1_N. 
 \label{schroe} 
\end{equation}

For the sake of simplicity, we focus on the fundamental problem of ground state calculation, 
i.e.~computing the lowest eigenvalue and eigenfunction. 
Casting this problem into a variational framework
\begin{align*}
 E^*  & = \min    \bigl\{ \langle \Phi, H\Phi \rangle : \; \langle \Phi, \Phi \rangle   =1 ,\;   \Phi \in \mathbb{H}^1_N \bigr\}, \\ 
 \Psi & = \mbox{argmin} \bigl\{ \langle \Phi, H\Phi \rangle : \; \langle \Phi, \Phi \rangle   =1 ,\;   \Phi \in \mathbb{H}^1_N \bigr\},
\end{align*}
the Ritz-Galerkin approximation is obtained by minimizing only over a finite-dimensional subspace
$\mathcal{V}_N^d \subset \mathbb{H}^1_N$.

\subsection{Tensor Product Spaces} 
\label{subsec:SchSecQuant.TPS}

We may start on the fundamentals introduced above,
and formulate everything in terms of (discrete) second quantization.
For this, we consider the finite-dimensional tensor product space 
\begin{equation*}
 \mathcal{H}^d = \bigotimes_{i=1}^d \mathbb{K}^2 , \qquad   \mathbb{K} = \mathbb{R}  , \mathbb{C}  . 
\end{equation*}
In many instances in quantum chemistry, we can easily use the real numbers, i.e.  $\mathbb{K} = \mathbb{R}$
instead of the complex ones from  $\mathbb{C}$.  
We use the canonical basis $\{ |0 \rangle :=  \mathbf{e}^0  ,  | 1 \rangle  :=  \mathbf{e}^1 \}$ 
of the vector space $\mathbb{K}^2 $, 
where $(\mathbf{e}^0)_\mu = \delta_{\mu,0} $, $(\mathbf{e}^1)_\mu = \delta_{\mu,1}$.
Therefore any $|\mathbf{u} \rangle \in \mathcal{H}^d $ can be represented by 
\begin{equation*}
 |\mathbf{u}\rangle  = \sum_{\mu_1 =0}^1 \ldots \sum_{\mu_d=0}^1 U (\mu_1 , \ldots , \mu_d ) \;
 \mathbf{e}^{\mu_1}  \otimes \cdots \otimes   \mathbf{e}^{\mu_d}    .
\end{equation*}
Using this basis, we can identify $ |  \mathbf{u}  \rangle  \simeq U \in \mathcal{H}^d  $, 
where $U$ is  simply a d-variate functions
\begin{equation*}
 (\mu_1,\ldots,\mu_d) \quad\longmapsto\quad U (\mu_1,\ldots,\mu_d) \in \mathbb{K}, \qquad 
 \mu_i = 0,1, \quad i=1,\ldots, d,
\end{equation*}
depending on discrete variables, usually called indices $\mu_i = 0,1 $.  
$\mathcal{H}^d$ is  equipped with the inner product 
\begin{equation*}
 \langle U,V \rangle := \sum_{\mu_1,\dots , \mu_d  \in \{ 0,1 \} }  
 \overline{U (\mu_1,\ldots,\mu_d )} V (\mu_1,\ldots,\mu_d )   ,
\end{equation*}
and the $\ell_2 $-norm $ \|U \| = \sqrt{\langle U , U \rangle}$. 


\subsection{Discretization and second quantization}
\label{subsec:SchSecQuant.DiscrSecQ}

Typically the finite dimensional  subspace $\mathcal{V}_N^d\subseteq \mathbb{H}^1_N$, mentioned at the end of section 2.1,
 can be defined by the $N$-fold antisymmetric tensor product of 
univariate spaces  $H^1(\mathbb{R}^3 \times 
\mathbb{Z}_2 , \mathbb{K} )$, where $\mathbb{K} = \mathbb{R} , \mathbb{C}$.  
These univariate spaces are defined  by 
choosing a complete ortho-normal one-particle basis set consisting of spin-orbtial functions  
\begin{equation*} 
 B^d := \bigl\{\varphi_p \mid p= 1 , \ldots , d \bigr\} \quad\subseteq\quad
 B   := \bigl\{\varphi_p \mid p \in \mathbb{N}\bigr\} \quad\subseteq\quad
 H^1\Bigl(\mathbb{R}^3 \times  
 \mathbb{Z}_2, \mathbb{K}\Bigr). 
\end{equation*}


Let us address some remarks about basis sets, since their choice has a tremendous influence on  the 
accuracy of the solution. Typically an orthogonal set of basis functions is computed by a preliminary  computational step.
After a fully convergent  Hartree Fock calculation, the $\varphi_p$, $p=1, \ldots, d$  are the  first $d$ eigenfunctions of the Fock operator. 
These basis functions are global functions, they are called \emph{canonical molecular orbitals}. Sometimes
  localized orbitals are used, or \emph{natural orbitals}, which are the eigenfunctions of the one-particle density matrix. 

Choosing $N$ distinct indices $1\leq p_1 < \ldots < p_N \leq d \in \mathbb{N}$ out of $\{ 1, \ldots ,d\}$ 
defines the subset $\{ p_1, \ldots, p_N \}$.
Let us decipher this choice by a binary string $\mu :=  ( \mu_1 , \ldots , \mu_d)$, 
where $\mu_i =1 $ if $i$ is contained in the set $\{ p_1, \ldots, p_N \}$, and $\mu_i=0$ otherwise.  
With this choice at hand,  we build  the  \emph{Slater determinant} $\Psi_{\mu} $
\begin{equation*} 
 \Psi_{\mu}  ( \mathbf{r}_1 , s_1 ; \ldots ; \mathbf{r}_N , s_N )  :=  \Psi_{[p_1,\ldots,p_N]} 
 ( \mathbf{r}_1 , s_1 ; \ldots ; \mathbf{r}_N , s_N )  :=
 \frac{1}{\sqrt{N!}} \det \big( \varphi_{p_\xi}  (\mathbf{r}_\zeta , s_\zeta ) \big)_{\xi,\zeta=1}^N   .
\end{equation*}
In other words 
$\mu_i = 0,1$ denotes the occupation number of the orbital function $\varphi_i$.
The subspace $ \mathcal{V}^d_N $, called the \emph{Full CI (Configuration Interaction) space},
is defined as the linear hull of all Slater determinants,
which can be built from the possible choices of $ N$-element subsets of $\{ 1, \ldots, d\}$. 
Obviously its dimension grows combinatorially, 
i.e.~$\dim \mathcal{V}^d_N =\binom{d}{N}$.    
Then the infinite set $\mathbb{B}_N := \{\Psi_{[p_1,\ldots,p_N]} \mid p_\xi < p_{\xi+1}  \}$ 
is an ortho-normal basis of the space $\mathbb{H}^1_N$, 
and the finite $\mathbb{B}^d_N := \{\Psi_{[p_1,\ldots,p_N]} \mid  1 \leq p_\xi < p_{\xi+1} \leq d  \}$ 
forms an ortho-normal basis of a finite dimensional subspace $ \mathcal{V}^d_N \subseteq \mathbb{H}^1_N $.
That is,
\begin{align*}
 \mathbb{B}^d_N := \bigl\{\Psi_{[p_1,\ldots,p_N]} \mid  1 \leq p_\xi < p_{\xi+1} \leq d  \bigr\} \quad&\subseteq\quad
 \mathbb{B}_N   := \bigl\{\Psi_{[p_1,\ldots,p_N]} \mid p_\xi < p_{\xi+1}  \bigr\} \quad\subseteq\quad
 \mathbb{H}^1_N,\\
\mathcal{V}^d_N :=\mbox{Span }  \mathbb{B}^d_N \quad&\subseteq\quad \mathbb{H}^1_N=\mbox{Span }\mathbb{B}_N.
\end{align*}
We can embed this space into a larger space.   
For $0 \leq M \leq d$, 
the ensemble of all Slater determinants with particle number $M$, i.e.~the number of electrons,
forms an orthonormal basis of an antisymmetric $M$-particle Full CI space 
$\mathcal{V}^d_M :=\mbox{Span }\{\Psi_{[p_1,\ldots,p_M]}  |  1 \leq p_1< \ldots < p_M \leq d\}.$
By taking the  direct sum $\mathcal{F}^d = \bigoplus_{M=0}^d \mathcal{V}_M^d$,
one defines the discrete \emph{Fock space} $\mathcal{F}^d$.
The full Fock space can be obtained by taking the limit for $d\to \infty$.  
Since we consider only finite dimensional approximation,
we do not intend  to understand in what sense this limit might be defined or not. 
We delineate how a binary encoding of the indices of basis functions of the discrete Fock space $\mathcal{F}^d$
may be used for the computation of Schr\"odinger-type equations with (anti-)symmetry constraints.
As introduced above, we index each basis function $\Psi_{[p_1 , \ldots , p_M]} = \Psi_{\nu}$ 
by a binary string $\nu = (\nu_1,\ldots,\nu_d)$ of length $d$. 
With the canonical basis  $\mathbf{e}^0 = (1, 0)^T$, $\mathbf{e}^1 = (0, 1)^T$, 
we define an isometric mapping $ \iota : \mathcal{F}^d \to \mathcal{H}^d $ by 
\begin{equation*}
 \iota:  \Psi_{[p_1 , \ldots , p_M]} \quad\longmapsto\quad  \mathbf{e}^{\nu_1} \otimes \ldots \otimes \mathbf{e}^{\nu_d}  \in  \mathcal{H}^d =  
\bigotimes_{i=1}^d \mathbb{K}^2,\qquad \mathbb{K} = \mathbb{R}, \mathbb{C},
\end{equation*}
and 
$\mathbf{e}^{\nu_1} \otimes \ldots \otimes \mathbf{e}^{\nu_d}  \in \mathcal{H}^d$ 
can be considered as the function $ U = \delta_{\mu , \nu}$, 
i.e.~$(\mu_1,\ldots,\mu_d) \mapsto U(\mu_1,\ldots,\mu_d) \in \mathbb{K}$,
where $U( \mu_1,\ldots,\mu_d) = 1$ iff $ (\mu_1,\ldots,\mu_d) = (\nu_1,\ldots,\nu_d)$ and zero otherwise.

The optimizer of the  energy functional restricted to the finite dimensional space $\mathcal{V}_N^d$
is the solution of the finite dimensional eigenvalue problem 
\begin{equation} \label{eq:des} 
  \Pi^d_N H \Psi = E \Psi, \quad
  \Psi \in \mathcal{V}_N^d, \quad
  \langle \Phi , H \Psi \rangle = E \langle \Phi , \Psi \rangle,\quad
  \text{for all} \; \Phi \in \mathcal{V}_N^d,
\end{equation}
where $\Pi^d_N : \mathbb{H}^1_N \to \mathcal{V}_N^d$ is the $L_2$-orthogonal projection onto $\mathcal{V}_N^d$, 
and $E=E_{0,d}$ is the lowest eigenvalue of this problem. 
With the basis $\Psi_{\mu}$ at hand, 
the minimizer can be obtained as the solution of the linear system
\begin{equation} \label{eq:les} 
 \mathbf{H} \mathbf{u} = E \mathbf{u}, \quad
 (\mathbf{H})_{\nu\mu} = \langle  \Psi_{\nu} ,  H \Psi_{\mu} \rangle, \quad  
 \Psi = \sum_{\mu } \mathbf{u}_{\mu} \Psi_{\mu}   .
\end{equation}
Error estimates of the approximation made above 
can be deduced from basic convergence theory of the Galerkin method, see e.g.~\cite{yserlect}.  
A major problem is that due to the combinatorial scaling of the complexity 
even the solution of the above discrete problem remains completely infeasible,
except for extremely small problems.

The solution of the  {\em discrete stationary $N$-electron Schr{\"o}dinger equation} $\Pi^d_N H\Psi = E \Psi$ 
is an element of the Fock space $\mathcal{F}^d$,
subject to the constraint that it is constructed solely from $N$-particle Slater determinants. 
Identifying $\mathbf{u}_{\mu} = U(\mu_1,\ldots,\mu_d)$
the approximate wave function can be expanded by  
\begin{equation*}
 \Psi = \sum_{\mu } U( \mu ) \Psi_{\mu} ,  \quad  
  \mu = ( \mu_1, \ldots , \mu_d) ,\quad
 \mu_i = 0,1   , \quad   i=1, \ldots , d  .
\end{equation*} 
$\Psi$ being an $N$-particle wave function in $\mathcal{V}^d_N$
is equivalent to $U$ being an eigenvector of the number operator $\mathbf{P} = \sum_{p=1}^d \mathbf{a}_p^{\dagger} \mathbf{a}_p$,
as defined below.
The approximate ground state calculation by the Ritz-Galerkin method (\ref{eq:des})
leads to a linear eigenvalue problem
\begin{equation}
 \mathbf{H} U = E U , \quad
 U \in \mathcal{H}^d \cap \mbox{Ker  } ( \mathbf{P} - N \mathbf{I} )  , 
\end{equation}
which by now is formulated in the binary Fock space $\mathcal{H}^d$. 
The well known Slater-Condon rules \cite{helgaker} can be reformulated by the following result.

\begin{theorem}
The Hamiltonian $\mathbf{H}: \mathcal{H}^d \to \mathcal{H}^d$ 
resp.~number operator  $\mathbf{P}$ on $\mathcal{H}^d$, are given by 
\begin{equation*}
  \mathbf{H} = \iota \circ ( \Pi_{\mathcal{V}_N} H ) \circ \iota^{\dagger}, \quad 
  \mathbf{P} = \iota \circ P \circ \iota^{\dagger}  . 
\end{equation*} 
Using 
\begin{align*} 
{A} := \begin{pmatrix}
0  & 1   \\
 0 &  0   
\end{pmatrix}, \quad
{A}^{\dagger} =
\begin{pmatrix}
0  & 0   \\
 1 &  0   
\end{pmatrix}, \quad   
{S} := \begin{pmatrix}
1  & 0   \\
 0 &  -1   
\end{pmatrix}, \quad
{I} := \begin{pmatrix}
1  & 0   \\
 0 &  1   
\end{pmatrix},
\end{align*}
and, indicating by ${A}_{(p)}$ that ${A}$ appears on the $p$-th position in the product,
\begin{equation*}
 \mathbf{a}_p:= {S} \otimes \ldots \otimes {S} \otimes {A}_{(p)} \otimes {I} \otimes \ldots \otimes {I},
\end{equation*}
we obtain in terms of binary annihilation and creation operators $  \mathbf{a}_{p}  ,  \mathbf{a}_{p}^{\dagger } $, 
that
\begin{equation}
 \mathbf{H} = \sum_{ p, q =1 }^{d}  h^q_p  \mathbf{a}_{p}^{\dagger }  \mathbf{a}_q + 
 \sum_{a,b,p,q=1}^{{d}} g_{p,q}^{a,b}  \mathbf{a}_a^{\dagger }  \mathbf{a}_b^{\dagger }  \mathbf{a}_p  \mathbf{a}_{q}, \qquad
 \mathbf{P} = \sum_{ p =1}^{d}  \mathbf{a}_{p}^{\dagger }  \mathbf{a}_p.
 \label{eq:constrain}
\end{equation}
Here for $h = - \frac{1}{2}  \Delta + V_\text{ext}$,
with  exterior potential is
$V_\text{ext} = - \sum_{\eta=1}^K\frac {Z_\eta}{| \mathbf{r} - \mathbf{R}_\nu|}$,  
the coefficients  
\begin{equation*} 
 h^q_p  =  \langle q | h | p \rangle : = \langle \varphi_q , h  \varphi_p \rangle 
  = \sum_{s = \pm \frac{1}{2} } \int_{\mathbb{R}^3} 
 \varphi_q^* (\mathbf{r}, s ) h \varphi_p (\mathbf{r} , s ) d \mathbf{r}
\end{equation*}
are the well known single-electron integrals, and 
\begin{equation*}
 g^{a,b}_{p,q} =  \sum_{s,s' = \pm \frac{1}{2}} \int
\int \frac{ \varphi_a^* ( \mathbf{r}, s) \varphi_b^* (\mathbf{r}',
s') \varphi_q ( \mathbf{r}, s ) \mathbf{\varphi}_p (\mathbf{r}', s'
)}{|\mathbf{r}-\mathbf{r}'|} d \mathbf{r} d \mathbf{r}' 
\end{equation*}
are the two-electron integrals. (Although we presently work with real numbers, we have included
the general complex valued definitions). 
With this, the discrete (Full CI) Schr{\"o}dinger equation can be cast into the binary variational form 
of finding $U \in \mathcal{H}^d$ such that
\begin{equation*}
 U =  \mbox{argmin }_{V \in \mathcal{H}^d}  
\bigl\{ \langle \mathbf{H}V, V  \rangle  \mid \langle V,V \rangle =1 ,\;  \mathbf{P}V = N V \}. 
\end{equation*}
\end{theorem}

Let us finally remark that the above formulation is nothing but the formulation
in terms of \emph{Second Quantization}.
Let us remark that the representation in the second quantization,
in the way described above, depends strongly  on the basis set.
Unitary transformations among the orbital basis functions will change the actual coefficients $h^p_q$, $g_{p,q}^{a,b}$. 

\section{Tensor Product Approximation}
\label{sec:TPApprox}

\subsection{Hierarchical Tensor Representation and Tree Tensor Networks}
\label{sec:TPApprox.Tree}

In multi-configuration theory 
one is typically looking for a best basis set $\{\varphi_i \mid i= 1, \ldots, d ,\; d \geq N\}$ 
of orbital functions of given size, 
which minimizes  the ground state energy. 
Or more precisely we are looking for subspaces 
\begin{equation*} 
 V_i = \mbox{ Span } \bigl\{ |\mathbf{e}_{x_i}\rangle  \mid x_i =0, \ldots , n_i -1  \bigr\},\quad   i= 1, \ldots, d    . 
\end{equation*}
This concept of subspace approximation can be used 
for an approximation of a tensor
in tensor product spaces
\begin{equation*}
| \mathbf{u} \rangle  = \sum_{x_1 = 0}^{n_1-1}  \ldots \sum_{x_d =0}^{n_d -1 }    U    (x_1, \ldots , x_d  )\;  | \mathbf{e}_{x_1} \rangle 
\otimes \cdots  \otimes | \mathbf{e} _{x_d}  \rangle \quad \in \quad
 \bigotimes_{i=1}^d  V_i :=  \bigotimes_{i=1}^d \mathbb{K}^{n_i}   .
\end{equation*}
If there is no ambiguity with respect to the basis vectors 
$\{ | \mathbf{e}_{x_i} \rangle  \mid  x_i =0, \ldots, n_i -1  \}$,
we can identify $| \mathbf{u} \rangle $ with the discrete function
\begin{equation*}
\Big( (x_1,\ldots,x_d) \mapsto U(x_1,\ldots,x_d) \Big) , \quad  
( x_1 , \ldots, x_d ) \in \{ 0 , \ldots , n_1 - 1 \} \times \cdots \times \{ 0 , \ldots , n_d -1 \}  \ . 
\end{equation*}

In the \emph{Tucker representation} or approximation,
one is looking for good or even optimal bases 
\begin{equation*}
       \bigl\{ | \mathbf{b}^i_{k_i} \rangle \mid k_i=1,\ldots,r_i \bigr\} 
\simeq \bigl\{ x_i \mapsto  b_i(k_i , x_i ) \mid k_i=1,\ldots,r_i \bigr\}  
\end{equation*}
of size $r_i \leq n_i$ in each coordinate direction $x_i$, $i=1,\ldots,d$, 
yielding  the representation (or approximation)
\begin{equation*}
| \mathbf{u}\rangle =   \sum_{k_1=1}^{r_1} \cdots \sum_{k_d=1}^{r_d}
  C(k_1, \ldots , k_d  ) | \mathbf{b}^1_{k_1} \;  \rangle \otimes \cdots \otimes | \mathbf{b}^d_{k_d} \rangle,
\end{equation*} 
or in terms of coefficients 
\begin{equation} \label{eq:tucker} 
 U(x_1, \ldots , x_d )  = \sum_{k_1=1}^{r_1} \cdots \sum_{k_d=1}^{r_d}
 C(k_1, \ldots , k_d ) b_1 (k_1, x_1) \ldots b_d (k_d, x_d) .
 \end{equation}
%

However, this concept does not prevent exponential scaling in the numbers of degrees of freedom,
only $n_i$ is replaced by $r_i$. 
In particular, for $n_i=2$ the concept cannot be used without further improvements. 
The \emph{Hierarchical Tucker format} (HT) in the form introduced by \cite{hackbuschkuehn},  
extends the above idea of subspace approximation into a hierarchical or multi-level framework. 
This novel perspective has been proposed earlier in multi-configurational Hartree model (MCTDH) 
\cite{mayer} as well as in terms of  tree tensor network states \cite{schollwoeck}.
Following \cite{hackbusch}, we proceed in a hierarchical way. 
For the approximation of $U$,
we may need in  the partial tensor product space $ V_1 \otimes V_2 $ only a subspace $V_{\{ 1,2 \} } \subset  V_1 \otimes V_2 $ 
of  dimension $ r_{\{1,2\}}  \leq n_1  n_2$. 
Indeed $V_{\{ 1,2\}}$ is defined through a new basis 
\begin{equation*} 
 \bigl\{ |  \mathbf{b}^{\{ 1,2 \} }_{k_{ \{ 1,2 \}}} \rangle \mid  k_{ \{ 1,2\} }  = 1, \ldots , r_{\{ 1,2\} }  \bigr\}   , 
\end{equation*}
where the new basis vectors are given in the form 
\begin{equation*}
 | \mathbf{b}^{\{ 1,2 \} }_{k_{ \{ 1,2 \}}} \rangle = \sum_{x_1 =1 }^{n_1} \sum_{x_2 =1}^{n_2}
 U_{\{ 1,2 \} } ( k_{\{ 1,2 \} }  , x_1 , x_2 )  \; | \mathbf{e}_{x_1}  \rangle \otimes | \mathbf{e}_{x_2 } \rangle   . 
\end{equation*}
We  may continue, e.g.~by building a subspace  
$V_{\{ 1,2,3\} } \subset  V_{\{ 1,2\} } \otimes V_3  \subset V_1 \otimes V_2 \otimes V_3 $,
or $V_{\{ 1,2,3,4\} } \subset  V_{\{ 1,2\} } \otimes V_{\{ 3,4\}}$ and so on.  



This can be cast into the framework of a partition tree,
with leaves $\{1\}, \ldots \{d\} $, 
simply abbreviated here by $1,\ldots,d$, 
and vertices $\alpha \subset D:= \{1,\ldots,d\}$  
corresponding to the partition $\alpha = \alpha_\text{L} \cup \alpha_\text{R}$, 
e.g.~$\alpha  = \{ 1,2,3 \} = \alpha_\text{L} \cup \alpha_\text{R} = \{ 1,2\} \cup \{ 3\}$,
where $\alpha_\text{L} := \alpha_{\{ 1,2 \}}$ and $ \alpha_\text{R} : = \alpha_{\{ 3 \} } $. 
We call $\alpha_\text{L}$, $\alpha_\text{R}$ the sons of the father $\alpha$. 
In general we do not restrict the number of sons,
and define the \emph{coordination number} by the number of sons plus $1$ (for the father).  
Let $\alpha_\text{L},\alpha_\text{R} \subset D$ be the two sons of $\alpha\subset D$,  
then $ V_{\alpha } \subset V_{\alpha_\text{L}} \otimes V_{\alpha_\text{R}} $ has a basis defined by 
\begin{equation} \label{eq:transfer} 
 | \mathbf{b}^{\alpha}_{k_\alpha} \rangle  = \sum_{k_{\alpha_\text{L}}=1 }^{r_{\alpha_\text{L}} } \sum_{k_{\alpha_\text{R}}=1}^{r_{\alpha_\text{R}}} 
 U_{\alpha} ( k_\alpha,k_{\alpha_\text{L}},k_{\alpha_\text{R}} ) \, | \mathbf{b}^{\alpha_\text{L} }_{k_{\alpha_\text{L}}} \rangle \otimes | \mathbf{b}^{\alpha_\text{R}}_{k_{\alpha_\text{R}}} \rangle   .  
\end{equation}
The tensors $(k_\alpha,k_{\alpha_\text{L}},k_{\alpha_\text{R}}) \mapsto U_{\alpha} (k_\alpha,k_{\alpha_\text{L}},k_{\alpha_\text{R}})$ 
are called \emph{transfer} or \emph{component tensors}. 
The tensor $ U_{D} = U_{  \{ 1,  \ldots, d \} } $ 
is called the \emph{root tensor}.
Without loss of generality, all basis vectors, e.g.~$\{| \mathbf{b}^{\{ 1,2 \} }_{k_{ \{ 1,2 \}}} \rangle\}$,  
could be constructed to be orthonormal.
The tensor $U$ is completely defined by these transfer tensors.
It could be reconstructed by applying (\ref{eq:transfer}) recursively.
\begin{figure}
\centering
\includegraphics{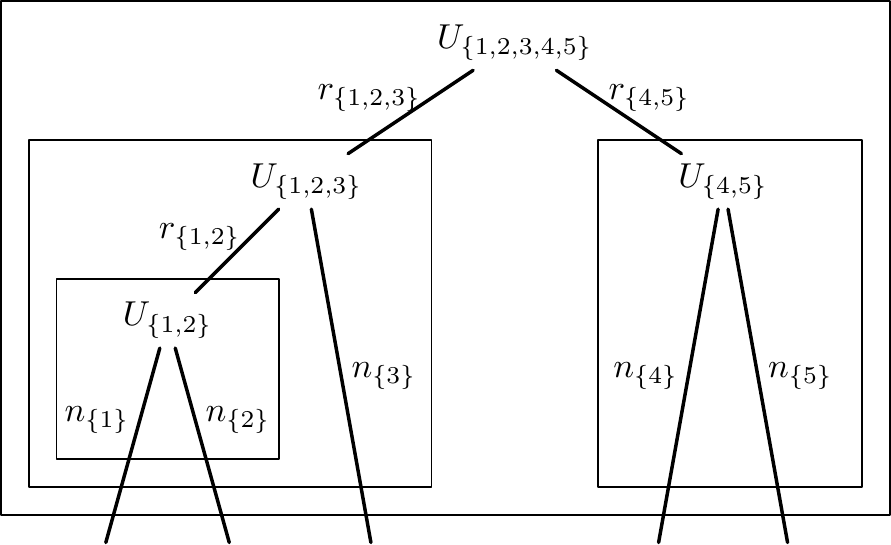}
\caption{Hierarchical Tensor representation}
\label{fig:HT}
\end{figure}

We highlight a particular case, namely \emph{matrix product states}, 
defined by taking $V_{ \{ 1, \ldots , i+1 \}} \subset V_{\{ 1, \ldots , i \} } \otimes V_{\{i+1\}} $. 
(Then we can abbreviate $\alpha = \{ 1,\dots,i\}$ simply by $\alpha:=i$, without any ambiguity.)
This form was developed as TT tensors (tensor trains) by \cite{osele,oselea} 
and turned out to be equivalent to matrix product states. 
The transfer tensors $U_{1,2,\dots,i} =:U_i$ are then of the form 
$\bigl( (k_{i-1} , x_i , k_i ) \mapsto  U_{i} (k_{i-1} , x_i , k_i )\bigr)  \in \mathbb{K}^{r_{i-1} \times n_i \times r_i }  $.  
Applying the recursive construction,  the tensor can be written by 
\begin{equation*}
 ( x_1, \ldots , x_d ) \quad\longmapsto\quad  U ( x_1, \ldots , x_d )  = \sum_{k_1=1}^{r_1} \ldots \sum_{k_{d-1}=1}^{r_{d-1}} 
 {U_1 ( x_1 , k_1) U_2(k_1,x_2,k_2) \ldots  
 U_d(k_{d-1},x_d )}  . 
\end{equation*}
If we introduce the matrices $\mathbf{U}_i (x_i) \in \mathbb{K}^{r_{i-1} \times r_{i} }$ by 
\begin{equation*}
 \big( \mathbf{U}_i (x_i) \big)_{k_{i-1},k_i } = U_i ( k_{i-1} , x_i , k_i )  ,\quad  1 < i < d 
\end{equation*}
together with the vectors
\begin{equation*}
 \big( \mathbf{U}_1 (x_1) \big)_{k_1 } = U_1 ( x_1 , k_1 )  ,\quad  \text{and}\quad
 \big( \mathbf{U}_d (x_d) \big)_{k_d } = U_d ( x_d , k_d )  ,
\end{equation*}
then we can represent the tensor by matrix products
\begin{equation*}
(x_1,\ldots,x_d) \quad\longmapsto\quad U(x_1,\ldots,x_d) = {\mathbf{U}_{1} (x_{1})  \cdots \mathbf{U}_{i} (x_{i}) \cdots \mathbf{U}_{d} (x_{d})}. 
\end{equation*} 
\begin{figure}
\centering
\includegraphics{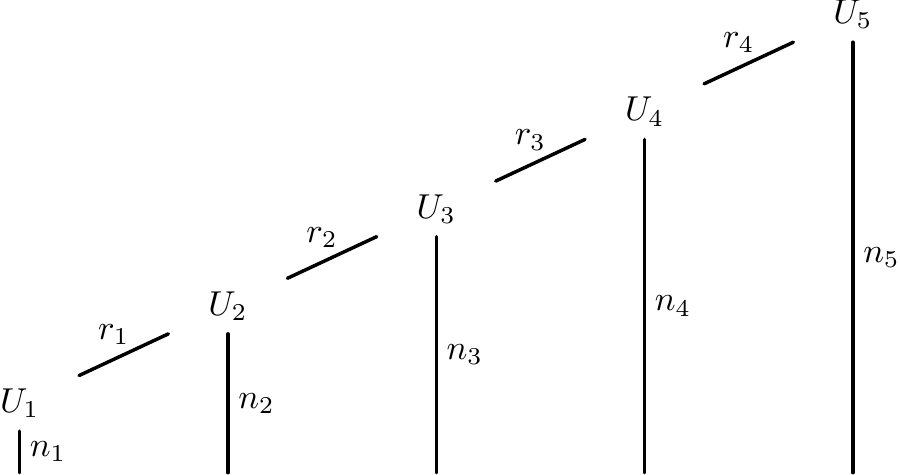}
\caption{Matrix Product State representation}
\label{fig:MPS}
\end{figure}

The tree is ordered according to the father-son relation in  a hierarchy of levels.
Using only orthogonal basis vectors, which is the preferred choice,
this ordering reflects left and right hand orthogonalization in matrix product states. 
We can rearrange the hierarchy in such a way that any $i =1 , \ldots , d$
can be  the root of the tree. 
Here $(k_{i-1} , x_i , k_i ) \mapsto  U_{i} (k_{i-1} , x_i , k_i )$ 
becomes  the root tensor. 
In the sequel we choose the matrix product states (TT format) as a prototype model for our explanations.
However, most properties can easily be extended to the general case with straightforward modifications.



The graphical representation in figure \ref{fig:HT} 
is an example of a tree tensor network state in quantum theory. 
Here the component tensors $ U_{\alpha} $ are called sites.
These are physical sites, if they contain at least one original variable $ x_{i} = \mu_i \in \{ 0, \ldots , n_i - 1\} $, 
otherwise they are considered as dummy sites. For  fermions, $ \mu_i = 0,1$ are occupation numbers $ n_i = 2$ and $U$ represents the state in the binary 
 Fock space $\mathcal{W} =: \mathcal{H}^d$.  
Each edge between sites denotes an index 
over which one has to perform a summation often called contraction.  
Removing an edge between two adjacent vertices will separate the original tree into two separate trees.
Roughly speaking it separates the full quantum system into two sub-systems. 
If $r_{\alpha} =1$, then this is a single tensor product (of pure states),
and separation will be perfect.
In this extreme case, we will call the systems to be disentangled. 
In general $r_{\alpha} >1$ is a measure how much these systems are entangled. 



The following result constitutes an important observation stemming from this separation. 
Let us consider only matrix product states (TT format) for simplicity, see e.g.~\cite{hrs-tt}. 
 
\begin{theorem}[Separation Theorem] 
For a given vertex $\alpha  :=  
 \{1 , \ldots , i  \}$ 
and $ D \backslash \alpha = \{ {i+1} ,  \dots , {d} \} $
the rank $ r_{ \alpha} = : r_i  $ 
is the separation or Schmidt rank of the matricization  $\mathbf{A}^i $ of $U( x_1, \ldots , x_d )$ 
casting the indices $(x_1 , \ldots , x_i )$ into a row index 
and the remaining ones $(x_{i+1} ,  \dots , x_{d} )$ into the column index of a matrix 
\begin{equation*}
 \mathbf{A}^{i}_{( x_1 , \ldots , x_i ), ( x_{i+1} ,  \ldots , x_{d} )} : = U ( x_1, \ldots , x_d )    .
\end{equation*} 
More precisely, we have the singular value decomposition
\begin{align*}
 \mathbf{A}^{i} & =   \mathbf{L} \mbox{ diag }  ( \sigma_{i,k} ) \mathbf{ R}^T   , \quad\text{or equivalently}   \\
 U ( x_1, \ldots, x_d) & =  \sum_{k_i =1}^{r_{i}}    L ( x_{1} , \ldots ,  x_{i } , k_i  )   \sigma_{i, k_i}  R(  x_{{i+1}} ,  \dots , x_{{d}} , k_i  )  . 
\end{align*}
\end{theorem}

The above theorem can be deduced from the \emph{high order SVD} (HOSVD) algorithm, 
which we explain for matrix product states known as Vidal decomposition \cite{vidal}, 
for HT see \cite{lars}.
This algorithm provides an exact reconstruction and can be used for approximation in a straightforward way, 
yielding quasi-optimal error bounds \cite{lars} for the corresponding approximation. 
\begin{enumerate}
 \item Given $U(x_1,\ldots,x_d)$, 
 \item matricization: $\mathbf{B}^1_{(x_1);(x_2,x_3,\ldots,x_d)} := U(x_1,\ldots,x_d) $;
 \item decomposition  (SVD): $ \mathbf{B}^1_{(x_1);(x_2,x_3,\ldots,x_d)} = \sum_{k_1=1}^{r_1} U_1(x_1,k_1)V_2(k_1,x_2,x_3,\ldots,x_d)$;
 \item \textbf{For} $i=2, \ldots, d-1$ \textbf{do}
 \begin{itemize}
  \item matricization: 
   $ \mathbf{B}^i_{(k_{i-1},x_i);(x_{i+1},\ldots,x_d)}:=V_i(k_{i-1},x_i,x_{i+1},\ldots,x_d)$,  
  \item decompose (SVD): $ \mathbf{B}^i_{(k_{i-1},x_i);(x_{i+1},\ldots,x_d)} = \sum_{k_i=1}^{r_i} U_i(k_{i-1},x_i,k_i)V_{i+1}(k_i,x_{i+1},\ldots) $;
 \end{itemize}
  \item $U_d(k_{d-1},x_d) := V_d(k_{d-1},x_d)$;
  \item $ U (\mathbf{x})  = \mathbf{U}_1 (x_1) \cdots \mathbf{U}_d (k_{d-1} , x_d) $.
\end{enumerate}

Remark: Let us consider the $\mathcal{H}^d = \bigotimes_{i=1}^d \mathbb{K}^2$ 
and $\| | \mathbf{u} \rangle  \| = \|U \|=1$,  
then  $(\mathbf{A}^{i})^* \mathbf{A}^{i}$ 
defines a density matrix at node $i$, 
with eigenvalues $ \lambda_{i, k_i} = \sigma_{i , k_i}^2 $. 
The decay behavior of the singular values $\sigma_{i,k_i}$ of $\mathbf{A}^i$ 
can be sharpened by introducing the \emph{block R{\'e}nyi entropy} 
of the density matrix $(\mathbf{A}^{i})^* \mathbf{A}^{i} $ with exponent $ \omega \in (0,1)$ 
\begin{equation*}
 S_{i}^{\omega}    := S^{\omega}\bigl((\mathbf{A}^{i})^* \mathbf{A}^{i}\bigr)
 :=  \frac{1}{1- \omega }  \log   \sum_{k_i =1}^{r_{i}} \sigma_{i , k_i }^{2\omega}  
 =  \frac{\omega}{1- \omega }  \log  \| ( \boldsymbol{\sigma}^2 )   \|_{\ell_{\omega}  }   ,
\end{equation*}
which is related to the \emph{von Schatten classes} where $2\omega = p$, given by 
\begin{equation*}
 \| \mathbf{A}^{i} \|_{*,p} := \| \boldsymbol{\sigma}_{i }  \|_{\ell_p}
 = \bigg( \sum_{k_i} \sigma_{i , k_i}^p  \bigg)^{\frac{1}{p} }  .
\end{equation*} 
In matrix product states these are called block entropies \cite{oers2}. 
From these entropies or \emph{von Schatten classes},
one can estimate the error of truncating the SVD at rank $r_i$. 
\begin{theorem}  \cite{uschmajew}
Let $ S_{i}^{\omega} $ be finite, with $\omega < 1$, for all $i$, 
i.e.~$\| | \mathbf{u} \rangle  \|_{*,2\omega} =  \sup_{i} \|\mathbf{A}^{i} \|_{*,2\omega }  < \infty $,
then $ | \mathbf{u} \rangle \in \mathcal{H}^d$ can  be approximated 
by a rank $\mathbf{r} = (\ldots , r_i , \ldots )$ tensor $ | \mathbf{u}_{\epsilon} \rangle$ 
with  an error bound
\begin{equation*}
 \||\mathbf{u} \rangle  - |\mathbf{u}_{\epsilon} \rangle \| \leq C  
 \big( \max \{ r_i \mid i=1,\ldots,d-1 \} \big) ^{\tau }   \sqrt{d}  \| |\mathbf{u} \rangle  \|_{*, 2 \omega },\quad
 \tau = \frac{1}{2 \omega }  - \frac{1}{2}  .
\end{equation*}
\end{theorem}

The multi-linear rank  $\mathbf{r} = (r_1, \ldots, r_{d-1} )$  
of a TT tensor is well well defined by the ranks $r_i$ of the matricisations $\mathbf{A}^i $.  
A tensor of given TT ranks $\mathbf{r} = (r_1, \ldots, r_{d-1} )$ 
can be reconstructed exactly  in MPS, resp.~TT format, by the Vidal decomposition described above, i.e.,  
by performing singular value decompositions over all matricisations $\mathbf{A}^i $.

\subsection{Hierarchical Tensors as Differentiable Manifolds} 
\label{subsec:TPApprox.Manif}

A central aim is to remove the redundancy in the parametrization 
of our admissible 
set $\mathcal{M}_\mathbf{r} {\subseteq\mathcal{H}^d}$,
which is the set of tensors of given TT rank $\mathbf{r}$.
(The situation becomes even more delicate when dealing with dynamical problems.)
Let us notice that, for example, the matrix product representation is not unique. 
In fact it is highly redundant. If we take 
a regular $r_1 \times r_1 $ matrix $ \mathbf{G}_1$,
we obtain by the following manipulation
\begin{equation*}
 U (\mathbf{x})  =  {\mathbf{U}_{1} (x_{1})  \mathbf{G}_1 \mathbf{G}_1^{-1} \mathbf{U}_{2} (x_{2})  \cdots \mathbf{U}_{i} (x_{i }) \cdots \mathbf{U}_{d} (x_{d})} 
 = { \widetilde{\mathbf{U}}_{1}  (x_{1})  \widetilde{\mathbf{U}}_{2} (x_{2})  \cdots \mathbf{U}_{i} (x_{i }) \cdots \mathbf{U}_{d} (x_{d})}
\end{equation*}
two different representations of the same tensor $U$.
Let us consider the space of parameters $(U_1 , \ldots , U_d)$, 
or $ \mathcal{U} :=  ( \mathbf{U}_1 (\cdot), \ldots , \mathbf{U}_d (\cdot))$,
together with  a (Lie)  group action. 
For regular matrices $\mathcal{G} =  (\mathbf{G}_1, \ldots , \mathbf{G}_{d-1} )$
this  group action is defined by 
\begin{equation*}
 \mathcal{G} \mathcal{U} := ( \mathbf{U}_1 (\cdot) \mathbf{G}_1, \mathbf{G}_1^{-1} \mathbf{U}_2 (\cdot) \mathbf{G}_2 , 
\ldots , \mathbf{G}_{d-1}^{ -1} \mathbf{U}_d(\cdot) )  .  
\end{equation*}
Having observed that tensor $U$ remains the same under this transformation of the component tensors,
we  identify two representations $\mathcal{U}_1 $ and $\mathcal{U}_2$,
if there exists $\mathcal{G}$ such that $\mathcal{U}_2 = \mathcal{G} \mathcal{U}_1$.
Standard differential geometry, similar to gauge theories in physics, 
asserts that this construction gives rise to a differentiable manifold  $ \mathcal{M}_{\mathbf{r}}$  \cite{lubich_blau,hrs-tt}.

The \emph{tangent space} $\mathcal{T}_U$ at $U \in \mathcal{M}_\mathbf{r}$,
i.e.~the space of all tangent directions,
can be computed from the Leibniz rule as follows.
A generic tensor $\delta U \in \mathcal{T}_U$ is of the form  
\begin{equation*}
\begin{split}
 \delta U ( x_1 , \ldots , x_d ) &=  E_1  ( x_1 , \ldots , x_d )  + \ldots + E_d (x_1, \ldots , x_d ) \\
 &=   \delta\mathbf{U}_1 (x_1) \mathbf{U}_2 (x_2) \cdots \mathbf{U}_d (x_d ) + \ldots  \\
  & \quad  + \cdots \mathbf{U}_{i-1} (x_{i-1} )  \delta \mathbf{U}_i  (x_i ) \mathbf{U}_{i+1} (x_{i+1} ) \cdots + \ldots + 
 \cdots \mathbf{U}_{d-1} (x_{d-1} )  \delta \mathbf{U}_d  (x_d)  . 
\end{split}
\end{equation*}
This tensor is uniquely determined if we impose \emph{gauge conditions}
onto $\delta U_i $, $i=1, \ldots , d-1$. 
Typically these conditions are 
\begin{equation}
 \sum_{k_{i-1} =1}^{r_{i-1}} \sum_{x_i =1}^{n_i} \overline{U_i (k_{i-1}, x_i , k_i ) }  \delta U_i (k_{i-1},x_i , k'_i ) =0, \quad
 \text{for all} \; k_i, k_i' = 1, \ldots , r_i  . 
\label{eq:gauge}
\end{equation}
We notice the following fact. 
For the root $d$ of the partition tree,
there is no gauge condition imposed onto $ \delta U_d$. 
The above gauge conditions (\ref{eq:gauge}) imply
that the $ E_i $ are pairwise orthogonal. 
Furthermore, the tensor $U$ is also included in the tangent space.
Curvature estimates are given in \cite{lrsv}. 

The manifold $\mathcal{M}_\mathbf{r} $ is an open set. 
It can  be  shown
that the closure of $\mathcal{M}_\mathbf{r}$ is $\mathcal{M}_{\leq \mathbf{r} }$,
the set of  all tensors with  ranks $r_i'$ at most ${r}_i $, $i=1,\ldots,d-1$. 
This is based on the observation that the matrix rank is an upper semi-continuous function \cite{hackbusch}.
The singular points are exactly those where the actual rank is not maximal.

\subsection{Dirac-Frenkel Variational Principle}
\label{subsec:TPApprox.DirFrenk}

We are going to approximate  the ground state by (multi-linear) rank $\mathbf{r}$ matrix product states,
by minimizing the energy expectation with respect to $N$ electron systems.
A natural setting would be to restrict to the set $\mathcal{M}_{\leq \mathbf{ r } }$,
but for technical reasons, let us consider 
the manifold $\mathcal{M}_\mathbf{r} $. 
 

Like for example in Hartree-Fock theory,
we to replace the original high-dimensional eigenvalue problem 
as a linear differential equation by much lower-dimensional,  but nonlinear equations.  
For the ground state calculation, we  would like to minimize the following energy functional
\begin{equation}
  \label{eq:ground-state} 
  \mathcal{E} (U) :=   \langle \mathbf{H} U , U \rangle  \quad\text{subordinated to}\quad 
  \| U \|^2   =  1  \quad\text{and}\quad  
  ( \mathbf{P} - N \mathbf{I} ) U   = 0  
  \quad\text{and}\quad
  U\in\mathcal{M}_\mathbf{r}    . 
\end{equation}
  
The first-order necessary condition for a minimizer of the problem (\ref{eq:ground-state})
can be formulated as follows, see e.g.~\cite{lrsv}.  
  
\begin{theorem}
If $U\in\mathcal{M}_\mathbf{r}$ is a minimizer of (\ref{eq:ground-state}) and 
$E=\langle \mathbf{H} U,U \rangle$, then
\begin{equation}
 \langle ( \mathbf{H} - E \mathbf{I} )  U , \delta U \rangle =0 , \quad  \text{for all}\;\delta U \in \mathcal{T}_{U}.
\label{eq:stationary} 
\end{equation}
\end{theorem}


Next,  we consider the dynamical problem
\begin{equation*}
 \label{eq:time-dependent}
  \frac{d}{dt} U =  \theta ( \mathbf{H} - E \mathbf{I} ) U, \quad 
   U (0)   =  U_0 \in \mathcal{M}_{\mathbf{r}}, 
\end{equation*}
where $\theta = -i $, $E=0$  corresponds to the time-dependent Schr\"odinger equation, 
and $\theta =-1$ corresponds to the gradient flow, often called imaginary time evolution.   
The Dirac-Frenkel variational principle \cite{lubich_blau}
requires that the approximate trajectory on a given manifold 
  $U_\mathbf{r}(t) \in \mathcal{M}_\mathbf{r} $   minimizes 
\begin{equation*}
 \| \frac{d}{dt} U  (t) -  \frac{d}{dt} U_\mathbf{r} (t)  \|  \to  \min, \quad U_{\mathbf{r}} (0) = U (0) . 
\end{equation*}
This leads to the weak formulation 
\begin{equation} \label{eq:dirac}
 \langle  \frac{d}{dt} U_\mathbf{r} - \theta (\mathbf{H} - E \mathbf{I} ) U_\mathbf{r} , \delta U \rangle = 0,\quad
 \text{for all} \; \delta U \in \mathcal{T}_{U_\mathbf{r}}  . 
\end{equation}

In the case that the manifold is simply a closed linear space
the equations above are simply the corresponding Galerkin equations. 
Let us further observe that in the static case, when $\frac{d}{dt} U=0$,
one obtains the first order condition (\ref{eq:stationary}). 
The Dirac-Frenkel principle is well-known in molecular quantum dynamics (MCTDH)
\cite{mayer,lubich_blau} for the Tucker format. 
For hierarchical tensors it has been formulated by \cite{mayer,wang}. 
First convergence results have been established recently  \cite{lrsv}. 
  

\subsection{DMRG and Alternating Linear Scheme} 
\label{subsec:TPApprox.DMRGALS}

We will demonstrate an efficient and fairly simple minimization method,
alled Alternating Linear Scheme (ALS),
which is based on the idea of alternating directional search. 
In contrast to poor convergence experienced with the canonical format (PARAFAC, CANDECOMP) \cite{kolda}, 
ALS implemented with some care in the hierarchical formats,
has been proved  to be 
surprisingly powerful.  
Furthermore, and quite important, it is robust against over-fitting,
i.e.~one can optimize in the set $\mathcal{M}_{\leq \mathbf{r}}$ \cite{hrs-als}. 
As a local optimization scheme, it converges only to a local minimum. 
This scheme is nothing but the one-site DMRG,
and could be improved by a modified version (MALS), 
which is the classical two-site DMRG algorithm \cite{white,schollwoeck}.
The basic idea of alternating direction gradient search 
is to fix all 
but only one component which is left to be optimized. 
Afterwards one turns to the next component repeating the procedure and iterate further. 
In tensor product approximation, 
this strategy was first used to find the best approximation, 
and called \emph{alternating least square} method. 
It is not surprising
that in each step one has to solve a small problem,
namely, to compute only a single component tensor $ \mathbf{U}_i (\cdot) $, 
resp.~$\bigl((k_{i-1} , x_i , k_i) \mapsto U_i (k_{i-1} , x_i , k_i)\bigr)
\in \mathcal{H}_i :=  \mathbb{K}^{r_{i-1} \times n_i \times r_i }$, (for fermions $n_i = 2$), 
when compared to the original problem in the full tensor space $\mathcal{H}^d$.  
Moreover, the smaller problem is of the same kind as the original problem. 
I.e.,~linear equations will be turned into small linear equations and eigenvalue problems 
will give rise to relatively small (generalized) eigenvalue problems.
In physics this supports the renormalization picture, 
where an original large system is reduced to a small system
with the same ground state energy, (and possibly further physical quantities). 
Due to the redundant representation of the components 
one cannot use the full parameter spaces $\mathcal{H}_i$,
but rather a nonlinear sub-manifold as shown below. 
But for the \emph{root} component there is no restriction.  
One can optimize in the full linear parameter space $ \mathcal{H}_i =  \mathbb{K}^{r_{i-1} \times n_i \times r_i }$. 
Before one moves on  to the next component, e.g.~${U}_{i+1} (\cdot)$,
one has to restructure the hierarchical tree 
to consider $ \mathbf{U}_{i+1} (\cdot) $ so as to be a root tensor.
In matrix product states 
this can be performed by left-hand (right-hand) orthogonalization of the formerly computed $ \mathbf{U}_i$.
The extension to general hierarchical trees is not simple, but straightforward. 
Since the Hamilton operator is the sum of tensor products of operators, 
we demonstrate the scheme only with a rank-one operator
$\mathbf{A} := \mathbf{A}_1 \otimes \cdots \otimes \mathbf{A}_d$.
The extension to the general case is easy,
as well as the generalization to Matrix Product Operators (MPO).

Given a (fixed) tensor $U^{(n)}$ in matrix product form $ U^{(n)} (\cdot) =: U (\cdot ) = \mathbf{U}_1 (\cdot) \cdots \mathbf{U}_d (\cdot) $,
let us consider the unknown component $ \mathbf{V} \in \mathcal{H}_i$, as being the root component.
We define a prolongation operator $\mathbf{E}_{i}$, by 
\begin{equation}
 \mathbf{E}_{i} : \mathcal{H}_i \longrightarrow \mathcal{H}^d,\quad
 \mathbf{E}_{i} \mathbf{V} ( \mathbf{x} ) =  \mathbf{U}_1 (x_1) \cdots  \mathbf{V} (x_i) \cdots \mathbf{U}_d (x_d)  ,
\end{equation}
which can be illustrated as
\begin{center}
\includegraphics{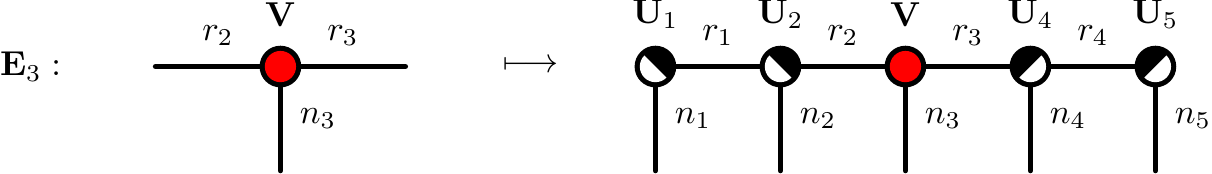}
 \\ALS ansatz
\end{center}
%
%
for example for $i=3$.

For solving the eigenvalue problem, formulated by a constraint optimization problem
\begin{equation*}
 U = \mbox{ argmin }  \bigl\{  \langle V,\mathbf{H}V \rangle \mid \langle V,V \rangle =1 ,\; V\in\mathcal{H}^d \bigr\},\quad
E_0 = \langle U , \mathbf{H} U \rangle,
\end{equation*}
we obtain a possibly  improved solution by solving 
\begin{equation*}
 \mathbf{U}_i^{(n)}  = \mbox{ argmin } \bigl\{  \langle  \mathbf{E}_{i}  \mathbf{V} , \mathbf{H} \mathbf{E}_{i} \mathbf{V} \rangle \mid
 \langle \mathbf{E}_{i} \mathbf{V}  , \mathbf{E}_{i} \mathbf{V} \rangle =1   ,  \mathbf{V} \in \mathcal{H}_i \bigr\} 
\end{equation*}
and 
\begin{equation*}
 U^{(n+1)} (\cdot) := \mathbf{U}_1 (\cdot) \cdots \mathbf{U}_i^{(n+1)}(\cdot) \cdots \mathbf{U}_d (\cdot). 
\end{equation*}
Let $\{\mathbf{e}_{k_{i-1},k_i}(x_i) \mid k_{i-1}=1,\ldots r_{i-1},\; k_i=1,\ldots,r_i,\; x_i=1,\ldots,n_i\}
$ be a basis, e.g canonical bases, of $\mathcal{H}_i$. 
Using a Lagrange-multiplier $E_0^{(n+1)} \in\mathbb{R}$,
the stationarity condition in weak formulation reads 
\begin{equation*}
\langle \mathbf{E}_{i} \mathbf{e}_i, \mathbf{H}\mathbf{E}_{i}\mathbf{V} \rangle  
- E_0^{(n+1)} \langle \mathbf{E}_{i} \mathbf{e}_i,\mathbf{E}_{i}\mathbf{V} \rangle = 0,\quad
\text{for all}\;\mathbf{e}_i \in \mathcal{H}_i. 
\end{equation*}
All possible contraction can be carried out,  and we obtain the problem
\begin{equation}
0 = \widetilde{\mathbf{H}}_i \mathbf{V} - E_0^{(n+1)} \widetilde{\mathbf{M}}_i  \mathbf{V} 
 =  \widetilde{\mathbf{H}}_i \mathbf{V} - E_0^{(n+1)} \mathbf{V}, 
\end{equation}
where $\widetilde{\mathbf{H}}_i  : \mathcal{H}_i \to \mathcal{H}_i $ is explicitly computable. 
Here, the matrix $\widetilde{\mathbf{M}}_i  = \mathbf{I} $ is the identity, 
due to the (right-hand and left-hand) orthogonality of the other components $\mathbf{U}_j$. 
Now, $ U^{(n+1)} \in \mathcal{H}^d$ may  be considered as an improved approximation of the 
ground-eigenstate, 
and $E_0^{(n+1)}$  approximates  the lowest eigenvalue $E_0 $ of $\mathbf{H}$. 

For a rank one operator  $\mathbf{A}$, the contracted operator is a matrix obtained by 
\begin{equation*}
\langle  \mathbf{E}_{i}  \mathbf{e}_{k_{i-1}, x_i , k_i }  ,
 \mathbf{A}  \mathbf{E}_{i} \mathbf{e}_{k_{i-1}', x_i' , k_i' }  \rangle 
   = \mathbf{L}^i_{k_{i-1} , k_{i-1}' } \otimes \mathbf{A}^i_{x_i , x_i' } \otimes \mathbf{R}^i_{k_i , k_i'}    ,   
\end{equation*}
where $\mathbf{e}_{k_{i-1}, x_i , k_i } , \mathbf{e}_{k_{i-1}', x_i' , k_i' } $
are orthonormal basis vectors in $\mathcal{H}_i$. 
For $i=1,\ldots,d $, the  left part and the right part could be computed recursively. 
Iterative solvers requiring only matrix-vector multiplications  
exploit  the tensor product structure and are preferred for the solution of the small systems. 


This ALS scheme, which is nothing but the  one-site DMRG algorithm, 
has the disadvantage that the ranks $r_i$ have to be
chosen a priori and cannot be increased  during this iteration procedure. 
In order to introduce higher ranks, 
one may do this in a greedy like fashion 
by adding to $ U^{(n)} $ a rank one (or rank $\mathbf{r}'$) tensor $V$, 
possibly chosen to be a best rank one (or rank $\mathbf{r}'$) surplus. 
The classical two-site DMRG or MALS is a more clever modification \cite{white}.
Instead of improving a single (root) component $ \mathbf{U}_i (x_i)$
one cast two adjacent components into one $ \mathbf{W}_i (x_i , x_{i+1} )$.
The enlarged parameter space is $\widetilde{\mathcal{H}}_i  = \mathbb{K}^{r_{i-1} \times n_i \times n_{i+1} \times r_{i+1} }$. 
In a decimation step the sought component $ U_i( \cdot) $ is computed from an SVD decomposition 
\begin{equation*}
{W}_i(k_{i-1},x_i,x_{i+1},k_i) \approx 
\sum_{k_i =1}^{r_i} U_i(k_{i-1},x_i,k_i) V_i(k_{i},x_{i+1},k_{i+1}).
\end{equation*}
Next turning to optimize $U_{i+1} (\cdot) $ one can use $ V_i(\cdot)$ as an initial guess. 
After optimizing $U_d (\cdot)$ one continues in reverse order and so on.


The correct orthogonalization or ordering in the tree 
provides the stability of this algorithm and its robustness with respect to over-fitting, 
since otherwise the density matrices would be singular. 
In \cite{hrs-als} is was shown that the corresponding condition numbers are bounded by  the condition number of the original operator 
\begin{equation*}
 \mbox{ cond } \widetilde{\mathbf{H}}_i \leq \mbox{Êcond } \widetilde{\mathbf{ H}_i} \leq \mbox{ cond } \mathbf{H}, 
\end{equation*}
and $E_0 \leq  E_0^{(n+1)} $ \cite{hrs-als}. 
It is obvious that the one-site DMRG is variational, 
but, due to the decimation step, 
the two.site DMRG is not exactly variational.   

\section{Coupled Cluster Method}
\label{sec:CC}


\subsection{Formulation of the Coupled Cluster  Ansatz}
\label{subsec:CC.Formulation}

Let us consider a \emph{reference determinant} $\Psi_0$,
which is usually the Hartree-Fock determinant, $\Psi_0 := \Psi_{[1,\ldots,N]}$.
Let us assume that this function 
is a good approximation to the exact ground state wave function $\Psi$. 
In practice, the complete basis $B$  
is substituted by a finite basis set $B^d$,
inducing a Galerkin basis $\mathbb{B}^d_N$
for a trial space (Full CI space) $\mathcal{V}_N^d$
contained in $\mathbb{H}^1_N$, as was recalled in section \ref{subsec:SchSecQuant.DiscrSecQ}.
We called $\{ \varphi_i  \mid i=1, \ldots, N\}$ \emph{occupied orbital functions}, 
since they are contained in the reference determinant. 
The remaining orbital functions $\{ \varphi_a  \mid a = N+1 , \ldots , d \}$
are called \emph{unoccupied}. 
For the construction and analysis of the CC method,
one can relax the orthogonality constraint, 
but it remains essential that 
\begin{equation*}
 \langle \varphi_i,\varphi_a \rangle = 0 \quad\text{for $i\leq N<a$}.
\end{equation*} 
If $ \Psi \not\perp \Psi_0 $,  
the solution $\Psi$ can be expressed as $\Psi =  \Psi_0 \oplus  \Psi^*$, 
i.e.~$\Psi^*$ is orthogonal to $\Psi_0$. 
Note that this $\Psi $ is not normalized by the $L_2$-norm,
but $ \langle  \Psi_0 , \Psi \rangle :=1$ provides the \emph{intermediate normalisation}.
Since the dimension of $\mathcal{V}_N^d$ grows combinatorially,        
$\mathbb{B}_d$ contains by far too many basis Slater determinants.
Therefore a subspace $\mathcal{V}_D$ of $\mathcal{V}_N^d$ 
might be chosen for discretisation. 
Mostly, the corresponding Galerkin method,
i.e.~the CI-ansatz loses \emph{size-consistency}. 
\emph{Size consistency} is an important issue emphasized by chemists.
It means that for a system $AB$ consisting of two independent subsystems $A$ and $B$,
the energy of $AB$ as computed by the truncated CI model 
is no longer the sum of the energies of $A$ and $B$.
This leads to inaccurate practical computations; 
therefore, the Full CI ansatz is replaced by a nonlinear ansatz \cite{cicehistory}, 
called the Coupled Cluster (CC) ansatz, which can easily be shown to be size-consistent \cite{helgaker,schneiderCC}.  

Let us fix a basis set according to the above requirements,
and turn to the binary  Fock space $ \mathcal{H}^d$.
The reference determinant $ \Psi_0$ corresponds to  the tensor  
\begin{equation*}
\iota (\Psi_0) =  \mathbf{c}_0 = \mathbf{e}_1^1 \otimes \cdots  \otimes \mathbf{e}_N^1 \otimes \mathbf{e}_{N+1}^0  
 \otimes \cdots  \otimes \mathbf{e}_d^0  
 \in \iota(\mathcal{V}_N^d)\subseteq \mathcal{H}^d .  
\end{equation*}
By second quantization the CC method is formulated in terms of \emph{excitation operators} 
\begin{equation*}
\mathbf{X}_{\beta} := 
\mathbf{X}_{i_1,\ldots,i_r}^{a_1,\ldots,a_r} : = \mathbf{a}^{\dagger}_{a_1} \cdots \mathbf{a}^{\dagger}_{a_r} 
\mathbf{a}_{i_1} \cdots \mathbf{a}_{i_r}   ,
\end{equation*}
where the excitation level is $r \leq N$,
and $i_1<\ldots<i_r\leq N$, 
and $N+1 \leq a_1<\ldots < a_r \leq d$, see \cite{helgaker}. 
There corresponds an excitation operator $X_{\beta }$ 
defined by their action on the basis functions 
$\Psi_{[p_1,\ldots,p_N]} \in \mathbb{B}^d_N$. 
If $\{p_1,\ldots,p_N\}$ contains all indices $i_1,\ldots, i_r$,
the operator replaces them by the orbitals $a_1,\ldots,a_r$;
otherwise, we let $ X_{i_1,\ldots,i_r}^{a_1,\ldots,a_r} \Psi_{[p_1,\ldots,p_N]} = 0$.

Indexing the set of all non-trivial excitation operators by a set $\mathcal{I}^d_N$, 
the \emph{cluster operator} of a coefficient vector
$\mathbf{t} = (t_{\beta})_{\beta \in \mathcal{I}^d_N} \in \mathbb{K}^{|\mathcal{I}_N^d|}$ 
is defined as $\mathbf{T}(\mathbf{t}) = \sum_{\beta \in \mathcal{I}^d_N} t_{\beta} \mathbf{X}_{\beta}$.
Choosing a suitable coefficient space $\mathbb{V}^d_N \subseteq \mathbb{K}^{|\mathcal{I}_N^d|}$ 
reflecting the $\mathbb{H}^1_N$-regularity of the solution (see \cite{RS,rohwedder}),
it can be shown that there is a one-to-one correspondence between the sets \cite{RS,rohwedder}
\begin{equation*}
 \bigl\{{\Psi}_0 + {\Psi}^* \mid {\Psi}_0 \bot {\Psi}^*\in \mathcal{V}_N^d \bigr\}, \quad
 \bigl\{\mathbf{c}_0 + \mathbf{c}^* \mid \mathbf{c}_0 \bot \mathbf{c}^*\in \iota(\mathcal{V}_N^d)\bigr\}, \quad
 \bigl\{\mathbf{c}_0 + \mathbf{T}(\mathbf{t})\mathbf{c}_0 \mid \mathbf{t}\in \mathbb{V}^d_N\bigr\}  \quad\text{and}\quad  
 \bigl\{e^{\mathbf{T}(\mathbf{t})}\mathbf{c} _0 \mid \mathbf{t}\in \mathbb{V}^d_N\bigr\}.
\end{equation*}
The latter exponential representation of all possible solutions
is used to reformulate the Full CI equations as the set of \emph{unlinked Full CC equations}
for a coefficient vector $\mathbf{t} \in \mathbb{V}^d_N$,
\begin{equation*}
\langle \mathbf{c}_{\beta}, (\mathbf{H} - E) e^{\mathbf{T}(\mathbf{t})} \mathbf{c}_0 \rangle = 0,\quad
  \text{for all} \; \mathbf{c}_{\beta} , \beta \in \mathcal{I}^d_N  . 
\end{equation*}
where $\mathbf{c}_{\beta}=\mathbf{X}_{\beta}\mathbf{c}_{0} $.
Inserting $e^{-\mathbf{T}(\mathbf{t})}$ yields the equivalent \emph{linked Full CC equations} 
\begin{equation*}
\langle \mathbf{c}_{\beta}, e^{-\mathbf{T}} \mathbf{H} e^{ \mathbf{T}} \mathbf{c}_0  \rangle  = 0,\quad
 \text{for all} \; \beta \in \mathcal{I}^d_N, \quad 
E^*  =  \langle \mathbf{c}_0, \mathbf{H} e^{\mathbf{T }} \mathbf{c}_0 \rangle. 
\end{equation*} 
For an 
underlying one-particle basis $B^d$, 
both of these two sets of equations are equivalent to the Schr\"odinger equation 
resp.~the linear Full CI ansatz \cite{RS,schneiderCC} under the condition 
that the functions $\varphi_1,\ldots,\varphi_N$ span an invariant subspace of an elliptic operator on $\mathbb{H}^1_N$, 
e.g.~the shifted Fock operator \cite{RS,schneiderCC}. 
The important difference between the CI and the CC ansatz,
aside from other advantages \cite{helgaker,kutzelnigg},
is that if the (usually much too large) index set $\mathcal{I}^d_N$ 
is restricted to some subset $\mathcal{I}_D$, 
the CC energy maintains the property of size-consistency as explained above, 
see \cite{helgaker} for further information. 
This restriction provides a projection 
and corresponds to a Galerkin procedure for the nonlinear function 
\begin{equation}
 \mathbf{f} : \mathbb{V}^d_N  \longrightarrow  (\mathbb{V}^d_N)',\quad
 \mathbf{f} (\mathbf{t})  :=  
 \bigl( \langle \mathbf{c}_{\alpha}, e^{-\mathbf{T}(\mathbf{t})} \mathbf{H} e^{\mathbf{T}(\mathbf{t})} \mathbf{c}_0 \rangle \bigr)_{\alpha \in \mathcal{I}^d_N}
\label{CCfunc}
\end{equation}
(for the linked case more suitable in practice, see below),
the solutions $\mathbf{t}^*$ of which correspond to 
solutions $e^{\mathbf{T}(\mathbf{t})}\mathbf{c}_0$ of the original Schr\"odinger equation. 
This gives the projected CC equations
\begin{equation*} 
 \langle \mathbf{f} (\mathbf{t}_D), \mathbf{v}_D\rangle  =  0,\quad
 \text{for all} \; \mathbf{v}_D \in \mathbb{V}_D, 
\end{equation*}
where $\mathbb{V}_D = \mbox{ Span } \{\mathbf{c}_{\beta}\mid\beta \in \mathcal{I}_D\}  \subseteq \iota(\mathcal{V}^d_N)\subset{\mathcal{H}^d}$ 
is the chosen Galerkin space,
indexed by a subset $\mathcal{I}_D$ of $ \mathcal{I}^d_N$,
i.e.~an equation for the Galerkin discretisation of the function $\mathbf{f}$:
\begin{equation}
\mathbf{f}(\mathbf{t}_D)   :=  \big( \langle \mathbf{c}_{\alpha}, e^{-\mathbf{T}(\mathbf{t}_D)} 
\mathbf{H} e^{\mathbf{T}(\mathbf{t}_D)} \mathbf{c}_0 
 \rangle \big)_{\alpha \in \mathcal{I}_D}  =  \mathbf{0}.
\label{CCdiscfunc}
\end{equation}
Usually, the Galerkin space $\mathbb{V}_D$ is chosen based on the so-called excitation level $r$ of the basis functions, 
i.e.~the number $r$ of one-electron functions 
in which $\Psi_{\beta}$ differs from the reference $\Psi_0$, 
see e.g.~\cite{helgaker},
or of pairs of creation and annihilation operators. 
For example, including at most twofold excitations ($r = 2$) 
gives the common CC Singles/Doubles (CCSD) method \cite{helgaker}.

\subsection{Numerical Treatment of the CC Equations}
\label{subsec:CC.Numerical}

It is common use to decompose the Hamiltonian into one- and two-body operators $\mathbf{H}= \mathbf{ F} +\mathbf{ U}$, 
where $\mathbf{F}$ is normally the Fock operator 
from the preliminary self-consistent Hartree-Fock. 
The one-particle basis set $\varphi_{p}$ consists of the eigenfunctions
as solutions of the discrete canonical Hartree-Fock equations with corresponding eigenvalues $\epsilon_{p}$. 
The CC equations (\ref{CCdiscfunc}) then read  
\begin{equation}
 \mathbf{F}_{\beta,\beta}t_{\beta} 
- \langle \mathbf{X}_{\beta}  \mathbf{c}_0, \sum_{n=0}^{4} \frac{1}{n!} [\mathbf{ U},\mathbf{T}]_{(n)} \mathbf{c}_0 \rangle = 0, \quad
  \text{for all} \;\beta \in  \mathcal{I}_D,
\end{equation}
with the \emph{Fock matrix} $ \mathbf{F} = \mbox{Êdiag } \big( \sum_{l=1}^r ( \epsilon_{a_{l} } - \epsilon_{i_{l}} ) \big) $. 
During the derivation of this equation it has been used that,
using the Baker-Campbell-Hausdorff formula 
and properties of the algebra of annihilation and creation operators \cite{helgaker},
for the Hamiltonian $\mathbf{H}$,
\begin{equation*}
 e^{-\mathbf{T}} \mathbf{H} e^{\mathbf{T}}  
=  \sum_{n=0}^{\infty} \frac{1}{n!} [\mathbf{H}, \mathbf{T}]_{(n)}  
=  \sum_{n=0}^4 \frac {1}{n!} [\mathbf{H}, \mathbf{T}]_{(n)},
\end{equation*}
with the $n$-fold commutators $[\mathbf{A} , \mathbf{B}]_{(0)} := \mathbf{A}$,
$[\mathbf{A} ,\mathbf{B} ]_{(1)} := \mathbf{A}\mathbf{B} - \mathbf{B}\mathbf{A} $, 
$[\mathbf{A} ,\mathbf{B} ]_{(n)} := [[\mathbf{A} ,\mathbf{B}]_{(n-1)}, \mathbf{B}]$.

The commutators are then evaluated within the framework of second quantisation,
resulting in an explicit representation of $\mathbf{f}$
as a fourth order polynomial in the coefficients $t_{\beta}$,
see \cite{crawschaeff} for a comprehensible derivation. 

The numerical treatment of the CC ansatz 
consists in the computation of a root of the nonlinear function $\mathbf{f}$.
This is usually done by application of quasi-Newton methods,
\begin{equation*}
 \mathbf{t}^{(n+1)}_D = \mathbf{t}^{(n)}_D - \mathbf{F}^{-1} \mathbf{f}(\mathbf{t}^{(n)}_D),
\end{equation*}
with an approximate Jacobian $\mathbf{F}$,
given by the diagonal Fock matrix,
provide that eigenfunctions of the Fock operator are used to constitute $B^d$.
On top of this method, it is standard to use the DIIS method 
(\emph{``direct inversion in the iterative subspace''}) for acceleration of convergence.

For the implementation of such a solver,
the discrete CC function (\ref{CCdiscfunc}) has to be evaluated.  
The most common variant of CC methods 
(often termed the ``Golden Standard of Quantum Chemistry'') 
is the CCSD(T) method,
in which first a CCSD method (see above) is converged and improved by a perturbational step.
While the computational cost for calculating small to medium sized molecules stays reasonable, 
it is thereby possible to obtain results that lie within the error bars of corresponding practical experiments \cite{bart_CC_overv,helgaker}.

\subsection{Lagrange Formulation and Gradients}
\label{subsec:CC.Lagrange}

A certain disadvantage of the CC  method is that it is not variational \cite{kutzelnigg}. 
E.g.~the computed CC energy is no longer a guarantied upper bound for the exact energy. 
The following duality concept can prevent from problems arising in this context.
Let us introduce a formal  Lagrangian
\begin{equation}
\label{lagr}
 L (\mathbf{t},\boldsymbol{\lambda} ): = 
 \langle \mathbf{c}_0, \mathbf{H} e^{\mathbf{T} (\mathbf{t} )} \mathbf{c}_0 \rangle 
 + \sum_{\alpha} \lambda_{\alpha} 
 \langle \mathbf{X}_{\alpha} \mathbf{c}_0, e^{ -\mathbf{T}(\mathbf{t})}\mathbf{H} e^{\mathbf{T}(\mathbf{t})} \mathbf{c} _0 \rangle. 
\end{equation}
With this definition at hand the CC ground state is given by 
$E=\inf_{\mathbf{t} \in \mathbb{V}^d_N} \sup_{\boldsymbol{\lambda} \in \mathbb{V}^d_N} L(\mathbf{t},\boldsymbol{\lambda})$.
The corresponding stationary condition with respect to $t_{\beta}$ reads
\begin{equation}
 \frac{\partial L}{\partial t_{\beta}} (\mathbf{t} , \boldsymbol{\lambda} )  
= \langle \mathbf{c}_0, \mathbf{H} \mathbf{X}_{\beta} e^{\mathbf{T} (\mathbf{t} )} \mathbf{c}_0 \rangle  
+ \sum_{\alpha}\lambda_{\alpha} 
\langle \mathbf{X}_{\alpha} \mathbf{c}_0 , e^{-\mathbf{T} (\mathbf{t} )} [\mathbf{ H} , \mathbf{X}_{\beta} ] e^{\mathbf{T} (\mathbf{t} )} \mathbf{c}_0 \rangle 
= E'(\mathbf{t})  +  \langle \boldsymbol{\lambda},\mathbf{f}'(\mathbf{t})\rangle  =  0 
\label{CClagrangian} 
\end{equation}
for all $ \beta \in \mathcal{I}_D $, 
while the derivatives w.r.t.~$\lambda_{\beta}$ 
yield exactly the CC equations $\mathbf{f}(\mathbf{t}) = 0$
providing the exact CC ground state  $ \mathbf{c}  = e^{\mathbf{T}(\mathbf{t})} \mathbf{c}_0$. 
Afterwards, the Lagrange multiplier $\boldsymbol{\lambda}$
can be computed from equation (\ref{CClagrangian}). 
Introducing the states 
\begin{equation*} 
\widetilde{\mathbf{c}} := \widetilde{\mathbf{c}} (\mathbf{t} , \boldsymbol{\lambda} ) 
= \mathbf{c}_0 + \sum_{\alpha}\lambda_{\alpha} e^{-\mathbf{T}^{\dagger} (\mathbf{t} )} \mathbf{X}_{\alpha} \mathbf{c}_0
= e^{ -\mathbf{T}^{\dagger} (\mathbf{t} )}  \biggl( 1 + \sum_{\alpha } \lambda_{\alpha}   \mathbf{X}_{\alpha} \biggr) \mathbf{c}_0,\quad
 \Psi(\mathbf{t})=  e^{\mathbf{T} (\mathbf{t} )} \Psi_0 ,
\end{equation*} 
there holds 
$  L (\mathbf{t} , \boldsymbol{\lambda} ) = \langle \widetilde{\mathbf{c}}
     (\mathbf{t} , \boldsymbol{\lambda} ),  \mathbf{H}  \mathbf{c}    (\mathbf{t} ) \rangle $ 
together with  the duality $\langle \widetilde{\mathbf{c}} , \mathbf{c}  \rangle = 1 $. 
As an important consequence,
one can compute derivatives of energy with respect to certain parameters, e.g.~forces,
by the Hellman-Feynman theorem.
If the Hamiltonian depends on a parameter $\omega$, $\mathbf{H} = \mathbf{H} (\omega )$,
then $\partial_{\omega}E= \langle \widetilde{\mathbf{c}} ,  (\partial_{\omega} \mathbf{H})  \mathbf{c}  \rangle$
holds for the respective derivatives with respect to $\omega$.
The above Lagrangian has been introduced in quantum chemistry
and the formalism has been extended further, 
e.g.~in \cite{Pedersen} for a linear, size-consistent CC response theory.

\subsection{Theoretical Results: Convergence and Error Estimates}
\label{subsec:CC.Theo}

Recently, it has been shown in \cite{RS}
that if the reference function $\Psi_0$ is sufficiently close to an exact wave function $\Psi$
belonging to a non-degenerate ground state
and if ${\mathbb{V}_D}$ is sufficiently large,
the discrete CC equation (\ref{CCdiscfunc}) locally permits a unique solution $\mathbf{t}_D$. 
If the basis set size is increased,
the solutions corresponding to $\mathbf{t}_D$ converge quasi-optimally in the Sobolev $H^1$-norm 
towards a vector $\mathbf{t} \in \mathbb{V}$ parametrizing the exact wave function $\Psi $.  
The involved constant (and therefore the quality of approximation) 
depends on the gap between lowest and second lowest eigenvalue and on $\|\Psi_0 - \Psi\|_{\mathbb{H}^1_N}$. 
The above assumptions and restrictions mean 
that CC works well in the regime of dynamical or weak correlation,
which is in agreement with practical experience.

The error $|E(\mathbf{t}) - E(\mathbf{t}_D)|$ of a discrete ground state energy $E(\mathbf{t}_D)$ 
computed on $\mathbb{V}_{D}$ 
can be bounded using the Lagrangian approach 
from the accuracy of the solution of the corresponding dual problem. 
Denoting by $(\mathbf{t}, \boldsymbol{\lambda})$ 
the stationary points of the Lagrangian (\ref{lagr}) belonging to the full energy $E$,
and by $\mathbf{t}_D$
the solution of the corresponding discretized equation $\mathbf{f}(\mathbf{t}_D)=\mathbf{0},$ 
the error of the energy can be bounded by
\begin{equation*}
|E(\mathbf{t}) - E(\mathbf{t}_D) | \lesssim 
\bigl(  d (\mathbf{t},\mathbb{V}_{D})  +  d (\boldsymbol{\lambda},\mathbb{V}_{D}) \bigr)^2
\end{equation*}
and thus depends quadratically on the distance $d(\cdot,\cdot)$ of the approximation subspace
to the primal \emph{and} dual solutions $\mathbf{t}, \boldsymbol{\lambda}$ in $\mathbb{V}$.
Note that these estimates are generalizations of error bounds for variational methods,
which allow for error bounds depending solely on $d(\mathbf{t},\mathbb{V}_{D})^2$,and an improvement of the error 
estimates given in \cite{kutzelnigg}.
Roughly speaking, 
this shows that CC shares the favorable convergence behavior of the CI methods,
while being superior due to the size-consistency of the CC approximation.    

\section{Concluding Remarks}
\label{sec:Concl}

Since this article was intended more for a tutorial purpose,
we do not go into the details of various branches of recent research. 
Whenever we mention the Coupled Cluster method  
we mean single reference projected Coupled Cluster, which is the standard procedure. 
Other variants are not easily computable 
or, like multi-reference CC, an object of intense recent research. 
Here, we consider only matrix product states (MPS to TT), 
since the  DMRG algorithm is a numerical technique treating this tensor representation.
We neglect improvement by general tree tensor network states of hierarchical tensor representation, 
which is also a subject of recent research \cite{verstraete}.  

\begin{itemize}
\item \emph{Importance of a reference determinant}: 
In CC the reference determinant is of crucial importance.
The tensor product ansatz in MPS (DMRG) does not require a pronounced reference determinant. 
It provides a somehow controlable approximation of the Full CI wave function.
Therefore it seems to be well suited for multi-reference situations \cite{reiher}. 
It could be viewed as an improvement of multi reference methods as MCSCF, dealing with the Full CI part.  
\item \emph{Electron-electron cusp}: 
CC, and for example, the CCSD ansatz have the potential 
to describe the electron-electron cusp appropriately,
up to the remaining basis set error.  
In the MPS (TT tensors) the ranks appear quadratically in the complexity.
Therefore as a single particle factorization ansatz,
the electron-electron cusp is  only being approximated in a suboptimal way. 
\item \emph{Size consistency:} 
Both method are size consistent.
Where in DMRG the size consistency is only for certain separations.
But on the other hand it describes the separation precisely,
even when the subsystems are not independent. 
\item \emph{Entanglement:}
Moderate entanglement is crucial for the complexity of MPS and the  DMRG algorithm. 
\emph{Area laws} are only known for quantum lattice systems,
see e.g.~\cite{schollwoeck} for further references.
The multiplicative representation of the CC ansatz, e.g.~CCD,  
\begin{equation*} 
 \mathbf{c} = \prod_{i<j;a<b} \bigl(\mathbf{I} + t_{i,j}^{a,b} \mathbf{a}_a^{\dagger}\mathbf{a}_b^{\dagger}\mathbf{a}_j\mathbf{a}_i \bigr) \mathbf{c}_0,
\end{equation*} 
shows that CC can even represent some highly entangled states in a data sparse way,  
since it is a product of rank two operators.
\item \emph{Scaling:}
The matrix product states scales 
for storage as $ \mathcal{O} ( d r^2  + d^4 )$ and as $ \mathcal{O}(d^3 r^3)$ for computational work in DMRG.
CCSD resp. CCSDT are roughly scaling as $\mathcal{O} ( d^4 )$, resp.~$ \mathcal{O} ( d^6 )$, for storage 
and as $\mathcal{O} ( d^6 )$ resp.~$ \mathcal{O} ( d^8  )$ for computational work.
If we assume a scaling $r = \mathcal{O} (d^2)$, 
we may observe that DMRG and CCSDT seem to be of comparable cost. 
Low order scaling techniques  \cite{densityfitting} and further enhancements may
reduce the scaling exponent as well as the pre-factors. 
\end{itemize}


\end{document}